\documentclass[a4paper,nodate]{sab} 

\usepackage[english]{babel}
\usepackage{pgfplots}%, Allen2008_Dopita2005
\pgfplotsset{compat=1.18}
\usepackage{amsmath,amssymb}
\usepackage{graphicx} 
\usepackage{txfonts,natbib} 
\bibpunct{(}{)}{;}{a}{}{,} 
\usepackage{hyperref}
\usepackage{url}
\usepackage[T1]{fontenc} 
\usepackage{t1enc} 
\usepackage{dirtytalk}
\usepackage{tabulary}
\usepackage{booktabs,siunitx,threeparttable}
\newcolumntype{L}{>{\raggedright\arraybackslash}X}% <-- added
\usepackage{ltablex}% <-- added
\usepackage{siunitx}% <-- added
\usepackage{caption}% <-- added
\renewcommand\ackname{Acknowledgements.} 
\renewcommand\andname{\&}
\renewcommand\keywordname{Keywords.} 
\renewcommand\figureptname{Figure}
\renewcommand\tableptname{Table}

\renewcommand\noteaddname{Note added to proofs.}
\renewcommand\andname{\&}

\idline{Boletim da Sociedade Astron\^omica Brasileira, {\bf 31}, no. 1, XX-XX}{3}
\begin{document} 
\title{The Local Group Symbiotic Star Population and its Weak Relation with Type Ia Supernovae} 
%\subtitle{IF YOU DO NOT HAVE A SUBTITLE, DELETE THIS LINE} 

\author{Marco Laversveiler\inst{1} \and Denise R. Gonçalves\inst{1}}
\institute{Observatório do Valongo -- Universidade Federal do Rio de Janeiro.\\\email{marcoaurelio18@ov.ufrj.br}}
\date{Received } 

\Abstract{Here we study the symbiotic stars (SySt) population and its relation with type Ia supernovae (SNe Ia) in the galaxies of the Local Group. SySt are low- and/or intermediate-mass evolved binary systems where a white dwarf (WD) accretes mass from a giant star. A fraction of these WDs can become massive enough to reach the Chandrasekhar mass. Therefore, SySt have been considered as potential SNe Ia progenitors. Taking two approaches, one empirical and another statistical, we estimated the SySt population on the Galaxy as having a minimum value of $1.69\times10^3$ and a expected one of $3.23\times10^4$. For Local Group dwarfs galaxies, the computed SySt population ranges from 2 to 4 orders of magnitudes lower. Concerning the SNe Ia with SySt progenitors, our general result is that SySt are not the main SNe Ia progenitors. On the other hand, we still expect that about 0.5--8\% of the SNe Ia have symbiotic progenitors in the Milky Way, while the majority of the -- low-mass -- dwarfs galaxies did not experience a symbiotic type Ia supernova.}{Apresentamos aqui um estudo da população de estrelas simbióticas (SySt) e sua possível relação com supernovas do tipo Ia (SNe Ia) nas galáxias do Grupo Local. SySt são sistemas binários evoluídos constituidos de estrelas de massa baixa e/ou intermediária, onde uma anã branca (WD) acumula massa de uma estrela gigante. Uma fração dessas WDs pode se tornar massuda o suficiente para atingir a massa de Chandrasekar. Portanto, as SySt são consideradas como potenciais progenitores de SNe Ia. Tomando duas abordagens, uma empírica e outra estatística, derivamos a população SySt na Galáxia como tendo um valor mínimo de $1,69\times10^3$ e um valor esperado de $3,23\times10^4$. Para galáxias anãs do Grupo Local, os valores correspondentes são de 2 a 4 ordens de grandeza menores. Em relação às SNe Ia com SySt como progenitoras, obtemos, como resultado geral, que as SySt não são as principais progenitoras de SNe Ia. Por outro lado, ainda esperamos que cerca de 0.5--8\% das SNe~Ia tenham progenitores simbióticos na Via Láctea, enquanto a maioria das galáxias anãs -- de baixa massa -- não experimentou SNe~Ia provenientes de SySt.}


%\titlerunning{<tirar comentário e preencher, se título for muito longo>}
%\authorrunning{<tirar comentário e preencher, se lista de autor for muito longa. Use o formato "I. G. Silva et al.">}

\authorrunning{Laversveiler, M. \& Gonçalves, D. R.}
\titlerunning{The LG Symbiotic Star Population and their Weak Relation with SNe Ia}
\maketitle 

\section{Introduction}
%
%<preencha o corpo do arquivo, incluindo figuras, tabelas e equações, quando necessário
%

Binary star systems can evolve in different ways, which primarily depend on the stars' zero age main sequence (ZAMS) masses, their initial orbital separation and eccentricity \citep{benacquista_2013}. Since a small difference in mass can lead to differences in evolutionary timescales during the main sequence (MS), one of the stars in a binary system will evolve first. If one of the stars becomes a giant and fills its Roche lobe (Roche lobe overflow -- RLOF), it will cause a flow of matter from the so-called donor star to its companion. The evolution of the RLOF binaries can be stable or unstable, depending on the donor's envelope structure (radiative or convective), and on the mass ratio of the system \citep{Ge_2010}. Stable systems will only experience a change in mass ratio, due to the mass flow and accretion on the companion. However, in unstable systems the feedback of the mass-loss on the effective Roche lobe radius ($R_L$) and on the envelope of the donor star leads to the disruption it's envelope and the engulfment of the companion, so the system enters a common envelope (CE) phase \citealp{paczynski_1976}; \citealp{Ge_2010}). The CE evolution can result in the merge of the stars or in the ejection of the envelope, producing a close evolved binary \citep{paczynski_1976}. In systems without RLOF, the mass transfer is limited to a fraction, due to stellar winds, and the components evolve more likely as if they were single star systems.

Binary stellar evolution can lead to the formation of symbiotic stars (SySt), which are evolved systems composed by low- and/or intermediate-mass stellar objects. In the typical configuration of SySt a white dwarf (WD) accretes matter from a red giant branch (RGB) or asymptotic giant branch (AGB) star, mostly via winds \citep{kenyon_2008}. However, there are evidences for SySt that have distorted giants, probably because of Roche lobe filling \citep{mikolajewska_2003}. Since SySt have accreting WDs, they have been considered as potential progenitors of type Ia supernovae (SNe Ia; e.g. \citealp{kenyon_1993}; \citealp{liu_2019}; \citealp{ilkiewicz_2019}). However, many SySt have low mass WDs, with RS Oph and T CrB being known exceptions \citep{mikolajewska_2011}, which is a counterargument regarding this type of SNe~Ia progenitor. Nevertheless, the fraction of SySt with massive enough ($\gtrsim 1.1$~M$_\odot$) WDs can be considered as promising progenitors of SNe Ia, contributing to the observed SNe Ia rate.

Our goal with this study is to characterize the binary systems with ZAMS properties compatible with the observed evolved of SySt. Then determine the evolutionary paths these systems could have taken, to reach their expected population in the Galaxy and in Local Group dwarf galaxies. This knowledge, combined with our statistical procedure, allows us to find the fraction of SySt with the minimum requirements to be considered progenitors of SNe Ia.

%\section{SySt Population}
\section{Lower Limit of Milky Way's SySt Population}

The lower limit in the Milky Way (MW) is obtained by studying the distribution of SySt, as a function of the Galactic height. Right ascension (RA), declination (Dec) and distance from \cite{akras_2019} and the updated online catalog of SySt \citep{merc_2019} were used to determine the distribution that better describes the data. After transforming the coordinates to the galactocentric ones, ($X_G$, $Y_G$, $Z_G$), and from a series of statistical tests -- maximization of log-likelihood, KS, and least squares -- we found that the best representation of the parametrized distribution is a Laplace distribution. From this distribution, we recover the scale height of the Galactic SySt as being $h = 0.654$ kpc.

The above derived parameters allow, via projection of the data on $Z_G = 0$ kpc, to compute the $1\sigma$ and $2\sigma$ data dispersion ellipses. Through the combination with the scale height of the disk, $H$, we computed the central SySt density of the MW, $n_0 \sim 1.0\text{--}2.7$~kpc$^{-3}$. This is the central value because it refers to the density at $Z_G = 0$~kpc. The lower limit for the SySt population, $N_\text{min}$, is then given by the integration of the distribution, scaled with $n_0$, in the Galaxy's volume, assuming cylindrical symmetry.

We used two values for the MW's disk radius, $R_\text{G}$. The first as four times the scale length of the thin disk, $R_\text{G} = 4h_d$, and the other as the truncation radius, $R_\text{G} = R_\text{trunc}$. Here, $h_d = 2.0\text{--}3.8$~kpc and $R_\text{trunc} = 16.1\pm1.3$~kpc are given by \cite{amores_2017}. With these values, we found $N_\text{min} \sim (1.2\text{--}2.8)\times10^3$, and for the best fit of $1.69 \times 10^3$ as the SySt population in the Galaxy.

%\subsection
\section{Statistical Binary Evolution and Expected SySt Population}

Our second approach to the problem of counting SySt in the LG relies on the use of observed properties of binary stars. Using them, we can statistically infer which evolutionary channel a given binary will follow in its evolution. From this approach, we can extract the population of SySt. To this end, three steps are needed: 1) finding the ZAMS physical characteristics a binary need to have to evolve to SySt; 2) statistically evolve this ZAMS fraction using pre-defined channels (e.g. \citealp{han_2020}; \citealp{lu_2006}); and 3) defining a parameter to scale the fraction computed with the expected population.

%\subsubsection{ZAMS Binary Physical Parameters}
\subsection{ZAMS Parameters}

Since the stars in SySt are evolved stars and of initial low- and/or intermediate-mass ($\sim 0.8\text{--}8.0$~M$_\odot$), we need to restrict our analysis to this subset of systems. The first important constraint is on the minimum mass, because both stars of the system need enough time to evolve into giant dimensions, in a timescale lower than the age of the Universe, which defines a threshold mass ($M_\text{thr}$). The values $M_\text{thr}=0.86\text{--}0.90$~M$_\odot$ are derived from the MS evolutionary timescale \citep{harwit_2006}, taking the reionization era epoch ($\sim13.3\times10^9$ yr; \citealp{schneider_2015}) as an upper boundary.

The second ZAMS parameter is a restriction on the mass ratio of the systems. It is defined as $q_\text{cut}(M_1) := M_\text{thr}/M_1$, being $M_1$ the primary mass, with mass ratio defined as $q := M_2/M_1$ ($M_2 \leq M_1$). This restriction is used to discard ZAMS binaries with $M_2 < M_\text{thr}$.

The third ZAMS parameter is the maximum orbital separation $a_\text{max}$. The latter is used to discard very wide binaries, which will basically evolve as the stars were singles. Kepler's third law, as a function of the primary mass ($M_1$) and mass ratio ($q$), give us the maximum orbital separation, $a_\text{max}(M_1,q)$, setting a maximum orbital period, $P_\text{max}$, as a fixed parameter. Given that this is a very uncertain parameter, we use a range of values $\log(P_\text{max}) \in [3.6,4.2]$ ($P_\text{max}$ in days), based on the largests orbital periods known for SySt (R Aqr: $\log(P) = 4.1$ -- \citealp{gromadzki_2009}; RR Tel: $\log(P) = 5.0$ -- \citealp{hinkle_2013}). Note that the orbital periods can increase or decrease during the system's evolution.

From Kroupa's initial mass function (IMF; \citealp{kroupa_2001}) for the primaries, $\xi(M_1)$ -- for which we assume that all systems are resolved --, a binary fraction, $f_\text{bin}(M_1)$ \citep{duchene_2013}, and the mass ratio and separation distributions ($\zeta(q)$ and $\zeta(a)$ -- \citealp{duchene_2013}), the fraction of ZAMS binaries with the desired physical characteristics is %then
\begin{equation}
    f_\text{bin}^* = \int_{M_\text{thr}}^8 \xi(M_1)f_\text{bin}\left[\int_{q_\text{cut}}^1\zeta(q)\left(\int_{a_\text{min}}^{a_\text{max}}\zeta(a)\:da\right)\:dq\right]\:dM_1.
\end{equation}

%\subsubsection{Statistical Binary Evolution}
\subsection{Binary Evolution Channels}

Having defined the initial population of binaries, we need to statistically consider their evolution. This is made considering three evolutionary channels:

\begin{itemize}
    \item[I.] The primary fills its Roche lobe during the MS phase;
    \item[II.] The primary fills its Roche lobe during the giant (RGB or AGB) phase;
    \item[III.] There is no RLOF during the entire evolution of the primary.
\end{itemize}

The selection of the channel for a given ZAMS binary is made by using the effective Roche lobe radius ($R_L$; \citealp{eggleton_1983})

\begin{equation}
    \frac{R_L}{a} \equiv x(q) = \frac{0.49\:q^{-2/3}}{0.6\:q^{-2/3} + \ln(1 + q^{-1/3})},
\end{equation}
and the expected radii of the primary ($R_\varphi$), which is computed as the temporal mean of the radius in a given evolutionary phase $\varphi$: MS, RGB, or AGB. The condition for RLOF is set as $R_\varphi(M_1) = R_L$, which then gives the restriction on the separation for each of the above channels as $a_{\text{cut},\varphi}(M_1,q) = R_\varphi(M_1)/x(q)$.

The RLOF in channels I and II can lead to a stable or unstable evolution, depending, strictly, on $M_1$ and $q$. The critical mass ratio, $q_\text{crit}(M_1)$, is computed in channel I based on \cite{ge_2013}, and on channel II based on \cite{chen_2008}, where we reconstruct $q_\text{crit}(M_1)$ with the assumptions made in this work. If $q \equiv M_2/M_1 > q_\text{crit}$ the system will have a stable RLOF, on the other hand, if $q < q_\text{crit}$ then the RLOF will be unstable.

The evolution through channel I can lead to a direct merger, a contact binary, or to a MS + He-WD (helium white dwarf) system, in the minority of cases. Since without a simulation we can't say exactly the fraction of MS + He-WD systems formed, we introduce a free parameter, $f_\ell^{\text{(I)}}$, to stand for this uncertainty. The idea is that the evolution of the secondary, for MS + He-WD systems, can lead to SySt ($f_\ell^{\text{(I)}}$ also takes it into account). Channel I gives the function $f_\text{evol}^{\text{(I)}}(M_1)$, which describes the fraction of ZAMS systems that becomes SySt, as a function of $M_1$, through the evolution described.

Evolution through channel II is divided in four sub-channels: RGB and AGB, stable or unstable; where RGB or AGB indicates that the ZAMS binary will fill its Roche lobe during RGB or AGB phases, and stable or unstable refers to their evolution during RLOF. Stable RLOF will lead to the formation of MS + WD binaries, while unstable RLOF will form a CE. The evolution through the CE phase is dynamic and results in the merge of the stars or in the ejection of the envelope. The outcome of CE ejection is a close MS + WD system. If CE is present, another free parameter, $f_\ell^{\text{(II)}}$, related to the fraction of systems that do not merge, is introduced. The WD in both scenarios (stable or unstable RLOF) can have a dominant composition of He or C+O, depending on the mass of the Roche lobe filling star and on its evolutionary phase. Analogously to channel I, this channel returns a function $f_\text{evol}^{\text{(II)}}(M_1)$, for each of its sub-channels.

Channel III is the simplest one. It takes into account the ZAMS binaries that do not experience RLOF during the evolution of the primary. It is only limited by the parameter $a_\text{max}$. Again, as channels I and II, this one returns the function $f_\text{evol}^{\text{(III)}}(M_1)$.

Finally, all channels are brought together to compute the fraction of SySt formed:
\begin{equation}
    f_\text{ss} = \int_{M_\text{thr}}^8\frac{df_\text{bin}^*(M_1)}{dM_1}\sum_i f_\text{evol}^{(i)}(M_1)\:dM_1,
\end{equation}
where the super index $i$ in $f_\text{evol}^{(i)}$ refers to the evolutionary channel it represents.

\subsection{The Scaling Parameter}

To obtain the SySt population, from the relative fraction derived above, we adopt the approach given by \cite{kenyon_1993}, which is based on formation rate of planetary nebulae (PNe). Thus, it is assumed that this rate closely represents the rate which stars with masses $> 0.6$~M$_\odot$ and $< 8$~M$_\odot$ complete their evolution. The scaling parameter is expressed as $\mathcal{N} = \mathcal{R}_\text{PN}\tau_\text{ss}$. Here $\mathcal{R}_\text{PN} = N_\text{PN}/\tau_\text{PN}$ is the formation rate of PNe and $\tau_\text{ss} \approx 5\times10^6$~yr \citep{kenyon_1993} is the timescale of the symbiotic phenomenon. For the Galaxy, we use this PNe rate as a density and combine it with the volume, $V$, of the disk, as $\mathcal{R}_\text{PN} = V\nu_{PN} = 2\pi R_\text{G}^2H\nu_\text{PN}$, where $\nu_\text{PN} \approx 2.4\times10^{-12}$~pc$^{-3}$~yr$^{-1}$ is the formation rate density of PNe \citep{phillips_1989}.

For the Local Group dwarf galaxies, we use a bolometric absolute magnitude approach for $\mathcal{R}_\text{PN}$. The PN population in a galaxy can be associated with the so called $\alpha$-ratio, which gives the number of PNe per unit bolometric luminosity of the galaxy \citep{buzzoni_2006}. Thus, the scaling parameter is given by
\begin{equation}
    \mathcal{N} = \mathcal{B}\tau_\text{ss}L_{\odot,\text{bol}}\times10^{0.4(M_{V,\odot} + \text{BC}_\odot - M_V - \text{BC})},
\end{equation}
being $\mathcal{B} \approx 1.8\times10^{-11}$ L$_{\odot,\text{bol}}^{-1}$ yr$^{-1}$ the specific evolutionary flux \citep{buzzoni_2006}, $M_{V,\odot} = 4.85$, $\text{BC}_\odot \approx -0.1$, $M_V$ the visual magnitude of the galaxies, and BC their bolometric correction ($-0.2$; \citealp{reid_2016}).

%\subsection{Results}

\subsection{Results for the Galaxy}

Since the majority of the multiple star systems are binaries, according to \citep{duchene_2013}, we adopt the binary fraction as the multiplicity frequency (MF); $f_\text{bin}(M_1) = \text{MF}(M_1)$. The MF gives the fraction of systems that are multiple, in this case as a function of the primary star's mass. We used the MF$(M_1)$ as given by \citep{duchene_2013}.

From an analysis of their impact on the final results, the free parameters were fixed to $f_\ell^{\text{(I)}} = 0.25$ and $f_\ell^{\text{(II)}} = 0.5$. In comparison with $f_\ell^{\text{(II)}}$ the parameter $f_\ell^{\text{(I)}}$ changes very little the expected number of SySt (up to few percents; $\sim 4\%$). The choice for $f_\ell^{\text{(II)}}$ is simply related to the difficulty in inferring a realistic value. As for $f_\ell^{\text{(I)}} = 0.25$ we just assumed a non-dominant fraction, since the formation of SySt through channel I is unlikely. The metallicity to compute the the stellar radii from models, per evolutionary phase, in the Galaxy, was an average of $Z = Z_\odot \approx 0.0134\text{--}0.0140$ from the models given by \cite{lagarde_2012} and \cite{claret_2019}.

\begin{figure}
    \captionsetup{size=tiny}
    \centering
    \includegraphics[scale = 0.21]{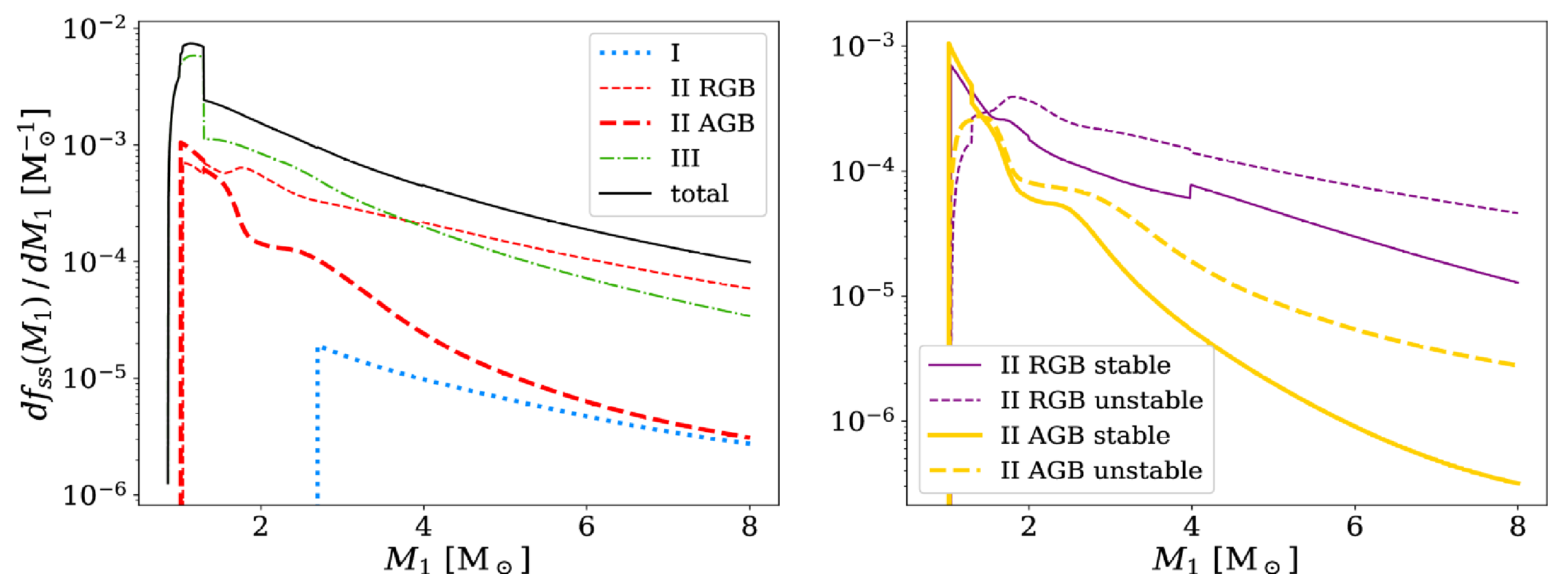}
    \caption{Example of the contribution from each evolutionary channel to $f_\text{ss}$ density (in mass space). In the left panel, we have: the blue doted line as channel I, the red thin dashed line as channel II RGB, the red thick solid line as channel II AGB, the green dash-doted line as channel III, and the black solid line as the sum of them all. On the right panel we have the contributions from each subset of channel II: yellow for the AGB and purple for the RGB channel; solid lines for the stable components and dashed for the unstable ones. For this plot, we used fixed $Z = Z_\odot$, $\log(P_\text{max}) = 4.2$, and $M_\text{thr} = 0.86$ M$_\odot$.}
    \label{df_ss_channels}
\end{figure}

\begin{figure}
    \captionsetup{size=tiny}
    \centering
    \includegraphics[scale = 0.35]{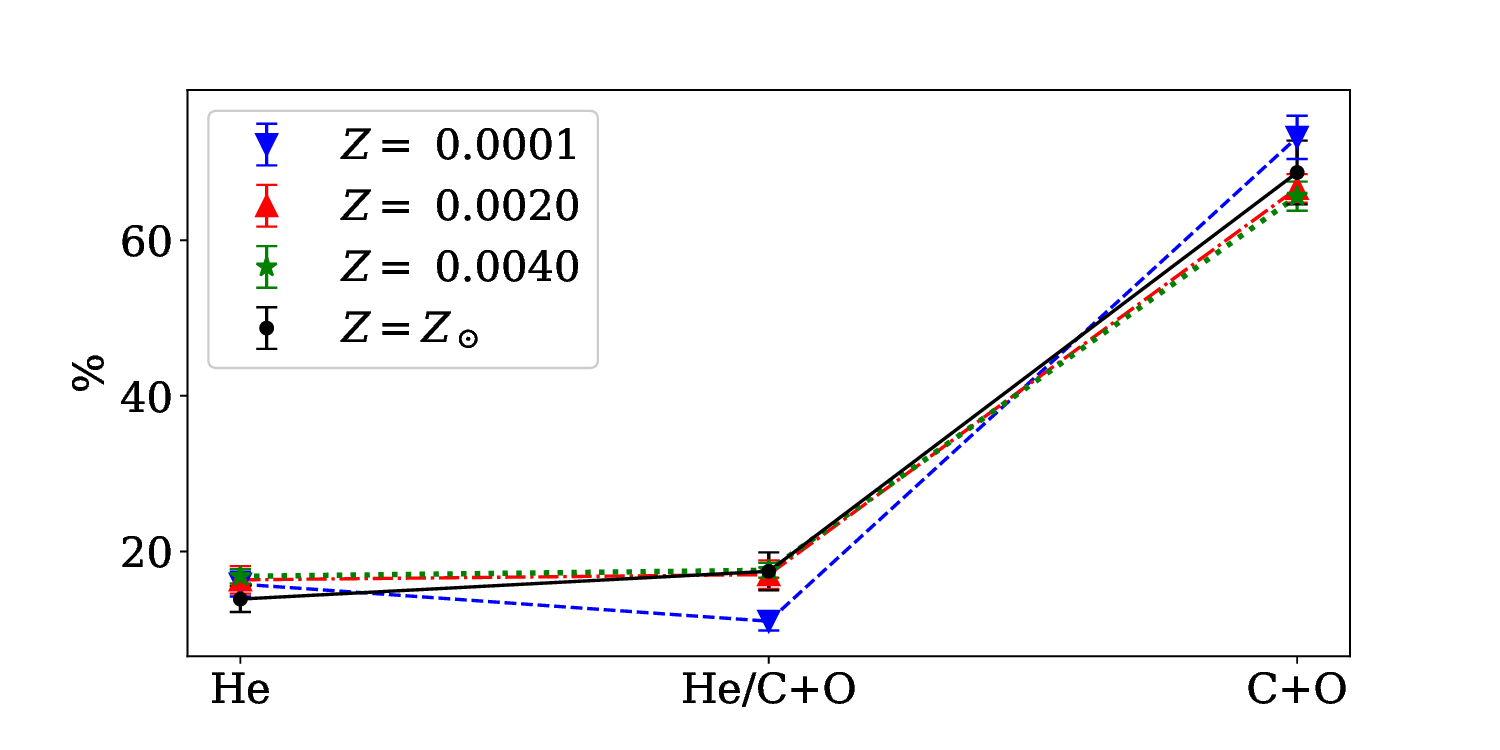}
    \caption{Composition of SySt's WDs obtained per metallicity model. The column He/C+O represents the percentage of SySt where we couldn't set an expected composition for the WD, or the composition is mixed}
    \label{WD_comp_results}
\end{figure}

Our main results for the Milky Way SySt are as in what follows: Figure \ref{df_ss_channels} displays the relative contribution per channel; Figure \ref{WD_comp_results} shows the expected chemical composition of the SySt's WDs found; and Table \ref{N_ss_result2} gives the expected SySt population.

As with the empirical approach, when the disk dimension is set to $R_\text{G} = R_\text{trunc}$, the resulting SySt population is an upper limit. When using $R_\text{G} = 4h_d$ we get the better fit, since it follows from the behavior of the galactic disk. Therefore, our best fit implies $3.23\times10^4$, with an upper limit of $6.18\times10^4$, SySt in the Galaxy.

\vspace{1cm}
Comparing our results with other authors' (e.g. $3\times10^5$ -- \citealp{munari_1992}; $3.3\times10^4$ -- \citealp{kenyon_1993}; $4\times10^5$ -- \citealp{magrini_2003}; $1.2\text{--}15.0\times10^3$ -- \citealp{lu_2006}), we note that it is in agreement with the previous estimations, and it is also very close to the value obtained by \cite{kenyon_1993}, when using $R_\text{G} = 4h_d$. This is not a coincidence, since we used an approach very similar to theirs in the computation of the scaling parameter. However, our stellar evolution considerations are more complex.

\begin{table}
    \captionsetup{size=tiny}
    \centering
    \caption[Galactic SySt population found given the different parameters]{Results for the Galactic SySt population, given the different parameters.}
    \label{N_ss_result2}
    \tiny{
        \begin{tabular}{c  c c  c c}
            \bottomrule
             & \multicolumn{2}{c}{$M_\text{thr} = $ 0.86 M$_\odot$} & \multicolumn{2}{c}{$M_\text{thr} = $ 0.90 M$_\odot$}\\
            \toprule
             & $R_\text{G} = 4h_d$ & $R_\text{G} = R_\text{trunc}$  & $R_\text{G} = 4h_d$ & $R_\text{G} = R_\text{trunc}$\\
            \midrule
            $\log(P_\text{max})$ & $N_\text{ss}$ & $N_\text{ss}$ & $N_\text{ss}$ & $N_\text{ss}$ \\
            $\log(\text{days})$ & $[\times10^4]$ & $[\times10^4]$ & $[\times10^4]$ & $[\times10^4]$\\
            \midrule
            3.6 & 3.02 $\pm$ 0.30 & 5.82 $\pm$ 0.98 & 2.76 $\pm$ 0.28 & 5.32 $\pm$ 0.90 \\
            3.9 & 3.38 $\pm$ 0.34 & 6.50 $\pm$ 1.10 & 3.07 $\pm$ 0.31 & 5.91 $\pm$ 1.00 \\
            4.1 & 3.63 $\pm$ 0.36 & 6.98 $\pm$ 1.18 & 3.29 $\pm$ 0.33 & 6.34 $\pm$ 1.07 \\
            4.2 & 3.75 $\pm$ 0.37 & 7.23 $\pm$ 1.22 & 3.40 $\pm$ 0.34 & 6.56 $\pm$ 1.11 \\
            \bottomrule
        \end{tabular}
    }
\end{table}

\subsection{Results for the Local Group Dwarf Galaxies}

We need to know the metallicity of the galaxies in order to study their $Z$-dependent characteristics. For that we use the converted [Fe/H] to $Z$, adopting $Z = Z_\odot10^{\text{[Fe/H]}}$ (see comment on table 6 of \citealp{mateo_1998}), and assign, per galaxy, the stellar evolution model with the closest $Z$. From \cite{lagarde_2012} we have the following metallicities available: $Z = 0.0001$; $Z = 0.0020$; $Z = 0.0040$. The IMF from \cite{kroupa_2001} and the mass ratio and separation distributions from \cite{duchene_2013} are also adopted here.

\begin{table}
    \centering
    \captionsetup{size=tiny}
    \caption[Parameters and results for Local Group dwarfs]{The results for the LG dwarf galaxies and the chosen parameters. Morphological type and metallicity are given by \cite{mcconnachie_2012}, $M_V$ by \cite{reid_2016}, and the known SySt population is given by the 2022 updated version of \cite{merc_2019} SySt catalogue.
    \tiny{
        \setlength\tabcolsep{0pt}
            \begin{tabular*}{\columnwidth}{@{\extracolsep{\fill}}rccccccccr}
                \toprule
                galaxy & type & $M_V$ & [Fe/H] & $Z$ used & $f_\text{bin}$ & $N_\text{ss}$ & known\\
                 & & [mag] & [dex] & [$\times10^{-3}$] & & & &\\
                \midrule
                LMC         & Ir   & $-$18.5 & $-$0.5  & 4.0  & 0.30$^\dagger$ & 685--849 & 10 \\
                SMC         & Ir   & $-$17.1 & $-$1.0  & 2.0  & 0.30$^\dagger$ & 187--232 & 12 \\
                %M 32        & E2   & $-$16,5 & $-$0,25 & 75,3 & D & (0,20--0,60)$^*$ & &   \\
                NGC 205     & Sph  & $-$16.4 & $-$0.8  & 2.0  & (0.25--0.55)$^*$ & 82--223 & 1  \\
                IC 10       & Ir   & $-$16.3 & $-$1.28 & 0.1   & (0.25--0.55)$^*$ & 75--203 & 1  \\
                NGC 6822    & dIr  & $-$16.0 & $-$1.0  & 2.0  & (0.25--0.55)$^*$ & 56--154 & 1  \\
                NCG 185     & Sph  & $-$15.6 & $-$1.3  & 0.1   & (0.25--0.55)$^*$ & 40--107 & 1  \\
                IC 1613     & dIr  & $-$15.3 & $-$1.6  & 0.1   & (0.25--0.55)$^*$ & 30--81 & 0 \\
                NCG 147     & Sph  & $-$15.1 & $-$1.1  & 2.0  & (0.25--0.55)$^*$ & 25--67 & 0 \\
                WLM         & dIr  & $-$14.4 & $-$1.27 & 0.1   & (0.25--0.55)$^*$ & 13--35 & 0  \\
                Sagittarius & dSph & $-$13.8 & $-$0.4  & 4.0  & (0.36--0.40)$^\ddagger$ & 11--15 & 0 \\
                Fornax      & dSph & $-$13.1 & $-$0.99 & 2.0  & 0.44$^\text{c}$--0.87$^\text{a}$ & 7--17 & 0  \\
                Leo II      & dSph & $-$10.1 & $-$1.62 & 0.1   & 0.33$^\text{b}$--0.36$^\text{a}$ & 0 & 0  \\
                Sculptor    & dSph & $-$9.8  & $-$1.68 & 0.1   & 0.58$^\text{a}$--0.59$^\text{c}$ & 0--1 & 0  \\
                Sextans     & dSph & $-$9.5  & $-$1.93 & 0.1   & 0.68$^\text{c}$--0.71$^\text{a}$ & 0--1 & 0  \\
                Carina      & dSph & $-$9.4  & $-$1.72 & 0.1   & 0.14$^\text{c}$--0.20$^\text{a}$ & 0 & 0  \\
                Draco       & dSph & $-$8.6  & $-$1.93 & 0.1   & 0.50$^\text{a}$ & 0 & 1  \\
                Ursa Minor  & dSph & $-$8.5  & $-$2.13 & 0.1   & 0.78$^\text{a}$ & 0 & 0  \\
                Hercules    & dSph & $-$6.6  & $-$2.41 & 0.1   & 0.47$^\text{d}$ & 0 & 0  \\
                Leo IV      & dSph & $-$5.5  & $-$2.54 & 0.1   & 0.47$^\text{d}$ & 0 & 0  \\
                \bottomrule
            \end{tabular*}
    }
    $^*$: Range of values used on absence of observed ones.\\
    $^\dagger$: Assumed based on the mean binary fraction value for LMC globular clusters \citep{milone_2009}, and in accordance with stellar formation history for SMC and LMC as commented by \cite{rubele_2011} and references therein.\\
    $^\ddagger$: Derived from radial velocity dispersion, but with high uncertainty \citep{bonidie_2022}.\\
    a: \cite{Spencer_2018}; b: \cite{Spencer_2017}; c: \cite{Minor_2013}; d: \cite{Geha_2013}.}
    \label{extragalactic_parameters}
\end{table}

Table \ref{extragalactic_parameters} contains the results obtained for the Local Group dwarf galaxies. We note that the expected value of the SySt population for this group is orders of magnitude lower than for the MW, which is expected, since, correspondingly, their masses are also orders of magnitude smaller. Moreover, from our analysis, the expected SySt population of a number of the LG dwarf galaxies is null. A way of interpreting these results is as an indicative that the formation rate of SySt, for the galaxies with $N_\text{ss} = 0$, is lower than the rate at which they cease to exist ($\sim 1/\tau_\text{ss}$). Draco is a good example of such an interpretation, since its SySt contradicts the expected value we obtained. For the remaining galaxies, the SySt population scales with their absolute magnitude in the V band, reaching a maximum of hundreds of SySt for the most luminous galaxies.

\cite{magrini_2003} also present results for the SySt population in some LG galaxies. However, they use an approach based on the galaxies' $K - B$ color to estimate their red giant population. Assuming that $0.5$\% of this population is in fact SySt. Their values are, in average, 100 times higher than ours. The discrepancy between their work and ours probably lies in the assumption of the $0.5$\% fraction, which can be interpreted as related with our $\mathcal{N}$ parameter. Again exposing the difficulty in finding a proper scale for the SySt population, with respect to the total stellar population of a galaxy.

\section{Can SySt be SNe Ia Progenitors?}

%\subsection{Method and Results}

This open question has been treated by a number of authors. A fairly good discussion with respect to this problem can be found, for example, in \cite{mikolajewska_2011}.

We can argue that WDs in SySt could accrete mass, from their giant companion, during the symbiotic phase, in a range of 0.05--0.25~M$_\odot$ \citep{mikolajewska_2011}. This implies that SySt with massive ($\gtrsim 1.1$~M$_\odot$) WDs could approach the Chandrasekhar mass limit ($M_\text{Ch} \approx 1.4$ M$_\odot$), then (possibly) experiencing nuclear instability and give rise to a SN~Ia event, according to the classic SNe~Ia model -- see \cite{hillebrandt_2000} for a review.

From our statistical binary evolution algorithm, with the lower WD mass limit of $M_\text{WD,min} = 1.1$ M$_\odot$, we fixed an upper limit to the fraction of SySt that could potentially be SNe~Ia progenitors ($f_\text{prog}$). We limit ourselves to channels II AGB and III, since they are the only ones that, certainly, will produce SySt with C+O WDs. We do not consider possible SNe Ia from common envelope phases during the evolution of the systems -- the core degenerate scenario \citep{soker_2019}. The integration limit $M_1(M_\text{WD,min})$ gives the minimum ZAMS mass of a star that generate a 1.1 M$_\odot$ WD, which was identified as $\sim 6.0$ M$_\odot$ using the initial-final mass relations (IFMR) given by \cite{cummings_2018} and references therein. Thus, we compute the rate of SNe~Ia with SySt progenitors as $r_\text{SNe Ia--SySt} = \mathcal{N}f_\text{prog}/\tau_\text{ss}$, and the respective timescale between consecutive SySt supernovae as $\Delta t_\text{exp} = 1/r_\text{SNe Ia--SySt}$. Table \ref{SNe_Ia_SySt} displays the results.

\begin{table}
    \centering
    \captionsetup{size=tiny}
    \caption[SNe Ia results]{SNe Ia results. The second column gives the computed fraction of potential SySt progenitors. The third, the formation rate of symbiotic supernovae. And the fourth, the timescale between each supernova event.}
        \label{SNe_Ia_SySt}
    \tiny{
        \begin{tabular}{c c c c}
            \toprule
            galaxy & $f_\text{prog}$ & $r_\text{SNe Ia--SySt}$ & $\Delta t_\text{exp}$ \\
                    & \%    & [yr$^{-1}$]            & [yr] \\
            \midrule
            Milky Way  & 1.22--1.70 & (7.37--24.6)$\times$10$^{-5}$ & (4.05--13.6)$\times$10$^{4}$ \\
            LMC         & 4.00--5.28 & (5.48--8.94)$\times$10$^{-6}$ & (1.11--1.82)$\times$10$^5$ \\
            SMC         & 4.02--5.29 & (1.50--2.46)$\times$10$^{-6}$ & (4.06--6.64)$\times$10$^5$ \\
            NGC 205     & 4.82--6.35 & (0.79--1.29)$\times$10$^{-6}$ & (7.74--12.6)$\times$10$^5$  \\
            IC 10       & 2.74--2.92 & (0.86--1.12)$\times$10$^{-6}$ & (8.89--9.93)$\times$10$^5$  \\
            NGC 6822    & 4.82--6.35 & (0.55--0.89)$\times$10$^{-6}$ & (1.12--1.83)$\times$10$^6$ \\
            NCG 185     & 2.74--2.92 & (4.52--5.90)$\times$10$^{-7}$ & (1.7--1.9)$\times$10$^6$ \\
            IC 1613     & 2.74--2.92 & (3.43--4.48)$\times$10$^{-7}$ & (2.2--2.5)$\times$10$^6$ \\
            NCG 147     & 4.82--6.35 & (2.38--3.90)$\times$10$^{-7}$ & (2.56--4.19)$\times$10$^6$ \\
            WLM         & 2.74--2.92 & (1.50--1.95)$\times$10$^{-7}$ & (5.1--5.7)$\times$10$^6$   \\
            Sagittarius & 3.34--3.96 & (7.22--11.90)$\times$10$^{-8}$ & (8.48--13.8)$\times$10$^{6}$  \\
            Fornax      & 1.82--2.74 & (3.78--6.17)$\times$10$^{-8}$ & (16.2--26.4)$\times$10$^{6}$  \\
            Leo II      & 4.19--4.35 & (2.85--3.72)$\times$10$^{-9}$ & (0.27--0.30)$\times$10$^9$  \\
            Sculptor    & 2.47--2.55 & (2.16--2.82)$\times$10$^{-9}$ & (0.35--0.39)$\times$10$^9$ \\
            Sextans     & 2.11--2.12 & (1.64--2.14)$\times$10$^{-9}$ & (0.46--0.52)$\times$10$^9$ \\
            Carina      & 7.53--10.25 & (1.50--1.95)$\times$10$^{-9}$ & (0.51--0.57)$\times$10$^9$  \\
            Draco       & 2.87--3.01 & (0.71--0.93)$\times$10$^{-9}$ & (1.19--1.07)$\times$10$^9$ \\
            Ursa Minor  & 1.84--1.93 & (0.65--0.85)$\times$10$^{-9}$ & (1.17--1.30)$\times$10$^9$ \\
            Hercules    & 3.05--3.21 & (0.11--0.15)$\times$10$^{-9}$ & (6.74--7.53)$\times$10$^9$ \\
            Leo IV      & 3.05--3.22 & (0.04--0.05)$\times$10$^{-9}$ & (18.6--20.7)$\times$10$^9$ \\
            \bottomrule
        \end{tabular}
    }
\end{table}

We note that $f_\text{prog}$ (Table~\ref{SNe_Ia_SySt}) is, on average, higher for the Local Group dwarf galaxies than for the Galaxy. This is easily explained by the use of $f_\text{bin}$ as a function of the primary mass for the Galaxy, but as a constant in the case of the LG dwarf galaxies. Since, in the Galaxy, the higher values MF are associated with the higher stellar mass population, which are the minority.

It is also interesting to note that the expected rate of SNe~Ia from SySt is very low, even for the Galaxy. Considering the total SNe~Ia rates estimated in the literature (e.g. $3\times10^{-3}$ yr$^{-1}$, \citealp{kenyon_1993}; $(5.4\pm1.2)\times10^{-3}$ yr$^{-1}$, \citealp{li_2011}; $14.1^{14.1}_{-8.0}\times10^{-3}$ yr$^{-1}$, \citealp{adams_2013}) we compute a contribution of about 0.5--8\% from SySt to the SNe~Ia rate. By comparing ours with the previous results, we conclude it is very unlikely that SySt are the main SNe~Ia progenitors. Nevertheless, SySt still cannot be ruled out as SNe~Ia progenitors in the classic SNe~Ia formation scenario, because a fraction of them will have massive enough accreting WDs (RS Oph and T CrB are well known examples -- \citealp{mikolajewska_2011}).

Regarding the result for the Local Group dwarf galaxies, there exists the possibility that some of them experienced a SNe~Ia from SySt during their evolution. At least in cases where $\Delta t_\text{exp}<10^{9}$ yr, since it is well restricted within the age of the Universe. For the remaining dwarf galaxies $\Delta t_\text{exp}$ is too high, and we conclude that no SNe~Ia from SySt has ever occurred on these galaxies.

\section{Conclusions}

This work is dedicated to the study of the population of symbiotic stars (SySt), with the goal of finding a robust way to estimate such population in the Milky Way and in the dwarf galaxies of the Local Group. Moreover, since SySt can satisfy the required characteristics for developing a SN~Ia event, we used our own algorithm to compute this specific fraction of SySt.

Using observational data, we adopted two approaches for the SySt population, one empirical and the other theoretical. We found that the SySt population in the Galaxy has a minimum value of $1.69\times10^3$, while its expected and upper limits are $3.23\times10^4$ and $6.18\times10^4$, respectively. For the dwarf galaxies, the value obtained ranged from zero to hundreds of SySt, which depended mostly on their bolometric absolute magnitude, with a weaker dependence on their metallicity.

Regarding the SNe~Ia, we obtained as a general result that SySt are not the main progenitors. Mostly due to the fact that the great majority of the WDs in SySt have masses below 1.1~M$_\odot$. This implies that the accretion rates in SySt are insufficient for them to reach the $M_\text{Ch}$. However, we found that a small fraction of the total SySt population could be progenitor of SNe Ia: in the Galaxy $\sim$ 1.5\%; and $\sim$ 3\% in the Local Group dwarf galaxies. By calculating the formation rate of SNe~Ia with SySt as progenitors, we show that 0.5--8.0\% of the SNe Ia in the Galaxy could come from SySt, and that most of the dwarf galaxies of the Local Group have not yet experienced SNe Ia from SySt.

%We presented here a study of SySt population and possible relation with SNe Ia in galaxies of the Local Group.

%Starting with an empirical approach to the galactic SySt population, using the observed distribution, we saw that we should expect to find at least $1.69\times10^3$ of them. The confirmed value today is 290 \citep{merc_2019}. We then decided to take a more theoretical approach, so we formulate an algorithm to statistically infer the number of expected SySt, in the Galaxy and Local Group dwarfs, based on pre-defined binary evolution channels and observed and simulated properties of binaries -- mass-ratio distribution, separation distribution, critical mass ratio etc. We got a best fit SySt population of $3.23\times10^4$ with an upper limit of $6.18\times10^4$ in the Galaxy. For Local Group dwarfs, the value was 2 to 4 orders of magnitude lower, with some cases ranging to zero SySt.

%About the SNe Ia with SySt progenitors, we got as a general result that SySt are probably not the main SNe Ia progenitors. However, they can't be ruled out, since, from our model, we should expect about 0.5--8\% of the SNe Ia to have symbiotic progenitors in the Galaxy. While the majority of the Local Group dwarfs probably did not experience any symbiotic supernovae.

%RETIRE A LINHA ABAIXO SE NÃO QUISER AGRADECER A ALGUÉM
%\begin{acknowledgements} ...  \end{acknowledgements} 
%

\begin{acknowledgements} We would like to thank Jaroslav Merc for providing us with the updated population of known SySt in the Local Group of galaxies. Authors acknowledge the following financial supports: ML, FAPERJ fellowship (2019); DGR, CNPq (313016/2020-8) and FAPERJ (Temático, 211-370/2021; CNE, 200.527/2023).\end{acknowledgements}

\end{document}